\documentclass[twocolumn,showpacs,preprintnumbers,amsmath,amssymb]{revtex4}
\usepackage{graphicx}

\begin{document}
\title{Relationship between quantum repeating devices and quantum seals}
\author{Guang Ping He}
 \email{hegp@mail.sysu.edu.cn}
\affiliation{School of Physics \& Engineering and Advanced Research
Center, Sun Yat-sen University, Guangzhou 510275, China}

\begin{abstract}
It is revealed that quantum repeating devices and quantum seals have
a very close relationship, thus the theory in one field can be
applied to the other. Consequently, it is shown that the fidelity
bounds and optimality of quantum repeating devices for decoding
quantum information can be violated when they are used for decoding
classical information from quantum states, and the security bounds
for protocols sealing quantum data exist.
\end{abstract}

 \maketitle

\newpage

\section{Introduction}

Suppose that many users share a single quantum communication channel
in multiuser transmission. Each user decodes the transmitted
information from the channel and passes the carrier to the
subsequent user. In this case, they need a quantum repeating device
\cite{qi586} in which the information can be decoded reliably, while
the quantum state of the carrier is expected to be optimally
preserved. (Note that such a device was called as \textquotedblleft
quantum repeater\textquotedblright\ in Ref. \cite{qi586}, but the
meaning differs from these in Refs. \cite{qi653,qi663}. To avoid
confusion, we prefer not to use the term quantum repeater in this
paper.) When quantum information is being transmitted, the
information-disturbance tradeoff of the devices was intensively
studied in literature \cite{qi639,qi637,qi638,qi636}. Optimal
quantum repeating devices were also proposed in Refs. \cite%
{qi586,qi587,qi666,qi667}. Nevertheless, in most cases of quantum
communication, a user generally cares little about the exact quantum
state of the carrier. Instead, he only wants to know the classical
information encoded in this quantum state. For example, in the
well-known quantum key distribution (QKD) problem
\cite{qi365,qi10,qi9}, a classical bit $0$ ($1$)
can be encoded as the quantum state $\left\vert 0\right\rangle $ or $%
\left\vert +\right\rangle $ ($\left\vert 1\right\rangle $ or
$\left\vert -\right\rangle $), where $\left\vert 0\right\rangle $
and $\left\vert 1\right\rangle $ are the two orthogonal states of a
qubit and $\left\vert \pm \right\rangle =(\left\vert 0\right\rangle
\pm \left\vert 1\right\rangle )/\sqrt{2}$. An eavesdropper is only
interested in whether the classical content is $0$ or $1$. There is
no need for him to distinguish $\left\vert 0\right\rangle $ from
$\left\vert +\right\rangle $ or $\left\vert 1\right\rangle $ from
$\left\vert -\right\rangle $. Therefore, it is more important to
study the tradeoff between classical information gained versus
quantum disturbance in quantum cryptography and search for optimal
quantum repeating devices for this purpose.

On the other hand, quantum seal (QS) is a relatively less-known
quantum cryptographic problem. Its goal can be summarized as
follows. The owner of the secret data to be sealed (denoted as
Alice) encodes the data with quantum states. Any reader (denoted as
Bob) can decode the data from these states without the help of
Alice. Meanwhile, if data has been decoded, it should cause a
disturbance on the states, which is detectable by Alice. A QS
protocol is considered to be secure if Bob cannot read the data
while escaping Alice's detection simultaneously. QS can be
classified by the types and the readability of the sealed data. If
the data are a single classical bit, it is called quantum bit seals
(QBSs). Else, if data are a classical string, it is called quantum
string seals (QSSs). If data can always be retrieved by the reader
with certainty, it is called a perfect QS. Else, if data can only be
retrieved with a non-vanished error rate, it is an imperfect QS. The
first perfect QBS protocol was proposed by Bechmann-Pasquinucci
\cite{sealing}, but then it was found that all perfect
QBS protocols are insecure against collective measurements \cite%
{impossibility}. Shortly later, it was proven that imperfect QBS
also has security bounds \cite{He}. Nevertheless, it was proven that
secure imperfect QSS protocols exist
\cite{String,Security,qi505,qi516}. (Note that it was claimed in
Ref. \cite{Insecure} that all QSS protocols are insecure. But as
indicated later in Ref. \cite{Security}, the cheating strategy
proposed in Ref. \cite{Insecure} is not a successful cheating
because it cannot obtain nontrivial amount of information while
escaping the detection
simultaneously \cite{Security,qi505}. It was also realized in Ref. \cite%
{qi516} that when the cheating in Ref. \cite{Insecure} escapes the
detection, the ratio between the amount of information obtained by the
cheater and that of the sealed string is arbitrarily small as the length of
the string increases.) More intriguingly, it was proposed in Ref. \cite%
{String} that secure QSS can be utilized to realize a kind of QBS, which is
secure in practice. Very recently, it was found \cite{PQBC} that QS has a
very close relationship with quantum bit commitment \cite{QBC}, which is
another primitive of quantum cryptography.

Though the theories of quantum repeating devices and quantum seals
are developed independently in literature, we can see that they are
closely related since they both focus on the information-disturbance
tradeoff on quantum systems. In the next section, we will show the
rigorous equivalence between their parameters. As the examples of
the application of this equivalence, we will apply the existing
theory of QS to study quantum repeating devices and obtain
interesting results on their fidelity bounds and optimality in Sec.
III. Also, we will apply the existing theory of quantum repeating
devices in Sec. IV, to study the security bounds for QS protocols
sealing quantum data.

\section{Equivalence between quantum repeating devices and quantum seals}

\subsection{Theory of quantum repeating devices}

Let us review briefly the description of quantum repeating devices
in Refs. \cite{qi586,qi587}. Suppose that an input state $\left\vert
\psi \right\rangle $ reaches a user via a quantum communication
channel. This user not only wants to decode the information of the
state for himself alone, but also wants to leave the state less
disturbed so that the subsequent user(s) can also decode some
information of the state without his help. For this purpose, he runs
a device which accomplishes the following tasks:

(i) A certain positive operator-valued measurement (POVM) $\{\Pi _{k}\}$ is
performed on $\left\vert \psi \right\rangle $.

(ii) When the outcome $k$ is observed at the output of the device, he uses
an inference rule $k\rightarrow \left\vert \phi _{k}\right\rangle $\ to
obtain the estimated signal state $\left\vert \phi _{k}\right\rangle $\ as
an approximation of the input state $\left\vert \psi \right\rangle $. Note
that the exact form of $\left\vert \phi _{k}\right\rangle $\ should be known
to the user. That is, e.g., when $\left\vert \phi _{k}\right\rangle $\ is a
qubit $x\left\vert 0\right\rangle +y\left\vert 1\right\rangle $, he should
know the values of $x$\ and $y$.

(iii) After the POVM, a conditional state $\left\vert \psi _{k}\right\rangle
$ (whose form depends on the value of $k$) is left to the subsequent user(s)
for further decoding.

Such a device is the quantum repeating device that we are referring
to. From this description, we can see that it is closely related
with the well-known quantum cloning machine \cite{qi121}, whose
purpose is also to transform an input state $\left\vert \psi
\right\rangle $ into two (or more) output states $\left\vert \phi
_{k}\right\rangle $\ and $\left\vert \psi _{k}\right\rangle $. The
difference is that at the end of the quantum cloning, the exact form
of $\left\vert \phi _{k}\right\rangle $ can still be
left unknown to the user. He may only own a quantum system whose state is $%
\left\vert \phi _{k}\right\rangle $. That is, a quantum repeating device can
be viewed as a quantum cloning machine plus a measurement on the output
state $\left\vert \phi _{k}\right\rangle $.

A user can choose the POVM at his will to construct his specific quantum
repeating device. Different choices will result in different output states $%
\left\vert \phi _{k}\right\rangle $\ and $\left\vert \psi _{k}\right\rangle $%
, which determine the quality of the device. Therefore it is natural
to seek for the choice which can optimize this quality. There are
two important parameters characterizing the quality of quantum
repeating devices, i.e., the transmission $F$ and estimation
fidelities $G$. Generally, we are interested in the case where the
possibility distribution of the input state looks completely random
to the user. In this case, the corresponding fidelities for the
given input signal $\left\vert \psi \right\rangle $,
averaging over all possible outcomes, are defined as \cite{qi586,qi587}%
\begin{equation}
F_{\psi }=\sum\limits_{k}p_{k}\left\vert \left\langle \psi \right.
\left\vert \psi _{k}\right\rangle \right\vert ^{2},
\end{equation}%
and%
\begin{equation}
G_{\psi }=\sum\limits_{k}p_{k}\left\vert \left\langle \psi \right.
\left\vert \phi _{k}\right\rangle \right\vert ^{2}.
\end{equation}%
Here, $p_{k}$ denotes the probability for the outcome $k$ to be
observed at the output of the quantum repeating device, so that the
input state $\left\vert \psi \right\rangle $ is decoded as the
estimated
signal state $\left\vert \phi _{k}\right\rangle $, while the state $%
\left\vert \psi _{k}\right\rangle $ is left to the subsequent user.
Performing average over all possible input states $\left\vert \psi
\right\rangle $, i.e., over the alphabet $A$ of transmittable
symbols (states), the transmission fidelity $F$ and the estimation
fidelity $G$ are given,
respectively, by%
\begin{equation}
F=\int\nolimits_{A}d\psi F_{\psi }
\end{equation}%
and%
\begin{equation}
G=\int\nolimits_{A}d\psi G_{\psi }.
\end{equation}

Some bounds on the values of $F$ and $G$ of different quantum
repeating devices were already found. Consider two extreme cases. In
the case where nothing is done by the quantum repeating device, the
input state is passed to the subsequent user unaltered and thus
$F=1$. Meanwhile, the outcome has to be estimated by guess, thus
$G=1/d$, where $d$ is the dimension of the Hilbert space of the
input states. In the opposite case where the quantum repeating
device gains the optimal information from the input state so that
the final state left to the subsequent user cannot provide any
information on the initial state, it was shown that $F=G=2/(d+1)$
\cite{qi637,qi638}.
Therefore we have%
\begin{equation}
2/(d+1)\leq F\leq 1  \label{qr1}
\end{equation}%
and%
\begin{equation}
1/d\leq G\leq 2/(d+1).  \label{qr2}
\end{equation}

In general cases where the quantum repeating device provides only partial
information while partially preserving the quantum state of the input signal
for the subsequent user, a tighter bound between $F$ and $G$ was found \cite%
{qi636} for randomly distributed input signals, i.e.,%
\begin{eqnarray}
&&(F-F_{0})^{2}+d^{2}(G-G_{0})+  \nonumber \\
&&2(d-2)(F-F_{0})(G-G_{0})  \nonumber \\
&\leq &(d-1)/(d+1)^{2},  \label{qr3}
\end{eqnarray}%
where $F_{0}=(d+2)/[2(d+1)]$ and $G_{0}=3/[2(d+1)]$. For two-dimensional
Hilbert space, the bound reduces to%
\begin{equation}
(F-2/3)^{2}+4(G-1/2)^{2}\leq 1/9.  \label{qr4}
\end{equation}

\subsection{Theory of quantum seals}

The model of QBS (i.e., the protocols sealing a single classical
bit) was established in Ref. \cite{He}. By analogy, here we
establish a general model of QS (i.e., covering the protocols
sealing any kind of classical bit(s), strings, or quantum
information) as follows:

(1) Alice, who owns the information $b$ to be sealed, maps $b$ into
a certain quantum state $\left\vert \phi \otimes \psi \right\rangle
$ of the system $\Phi \otimes \Psi $ and keeps the system $\Phi $\
to her own while making $\Psi $\ accessible by any potential reader
Bob who may want to decode $b$.

(2) Alice lets Bob know an operation $P$ for decoding. If the state of $\Psi
$\ is an eigenstate of $P$, the protocol is a perfect QS. Otherwise it is an
imperfect QS.

(3) Alice lets Bob know a series of sets $G_{i}$'s and a series of values $%
b_{i}$'s, which satisfies $G_{i}\cap G_{j}=\emptyset $ ($\forall i\neq j$),
such that if he applies $P$ on $\Psi $\ and the outcome $g$ belongs to set $%
G_{i}$, he should take the value of the sealed data as $b^{\prime
}=b_{i}$; while if $g$ does not belong to any $G_{i}$, the sealed
information cannot be identified, i.e., Bob needs to guess
$b^{\prime }$ by himself. (In the case where the sealed information
$b$ is a quantum data, each $b_{i}$ should be understood as a set of
parameters sufficient to describe a quantum state.) Note that since
the state of $\Psi $\ may not be an eigenstate of $P$, the value of
$b^{\prime }$ thus obtained will match Alice's input $b$ with a
certain probability only. Let $\alpha $ denote the average of this
probability over all possible $b$\ and $b^{\prime }$, which measures
the readability of the protocol.

(4) At any time, Alice can access to the entire system $\Phi \otimes
\Psi $ and compare its current final state $\left\vert \phi ^{\prime
}\otimes \psi ^{\prime }\right\rangle $ with the initial state
$\left\vert \phi \otimes \psi \right\rangle $. Therefore, if $b$ has
been read, Alice can detect it with the probability $1-\left\vert
\left\langle \phi \otimes \psi \right\vert \left. \phi ^{\prime
}\otimes \psi ^{\prime }\right\rangle \right\vert ^{2}$. Let $\beta
$ denote the average of this probability over all possible initial
and final states, which measures the security of the protocol.

For QBS, $b$ is limited to a single classical bit. In this case, it was
shown \cite{He} that the parameters $\alpha $\ and $\beta $\ in such a
protocol must satisfy the following security bounds:%
\begin{equation}
\beta \leq 1/2  \label{qs1}
\end{equation}%
and%
\begin{equation}
\alpha +\beta \leq 9/8.  \label{qs2}
\end{equation}

\subsection{Equivalence}

By comparing the above descriptions, we can see the relationship
between the two subjects. Suppose that Alice encodes a certain
information with a quantum state $\left\vert \psi \right\rangle $,
Bob decodes the information from $\left\vert \psi \right\rangle $
with a quantum repeating device suggested by Alice, and the
resultant state $\left\vert \psi _{k}\right\rangle $ is acquired
later by Alice to detect whether the information has been decoded.
Then the quantum repeating device in fact fulfills an embodied
implementation of quantum seals. On the contrary, Alice can use a
quantum seal protocol to encode some information on a quantum system
$\Psi $ and send it to Bob via the quantum communication channel,
Bob decodes it with the operation $P$ suggested by the protocol, and
another
subsequent user instead of Alice receives the final state of the system $%
\Psi $ for further decoding. Then the quantum seal works as a
quantum repeating device in this case. Therefore, quantum repeating
devices and quantum seals are in fact equivalent. That is, we can
set up a mapping between the elements of the two subjects as shown
in Fig. 1. Then a scheme for quantum repeating device can be
constructed from a quantum seal protocol and vice versa.

\begin{figure}
\includegraphics{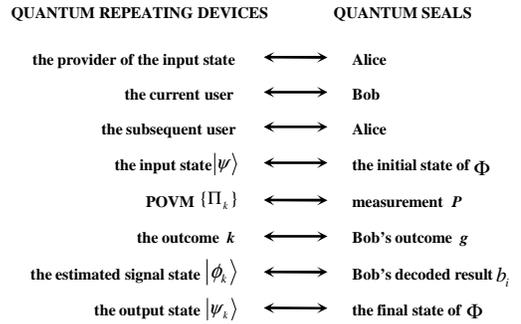}
\caption{\label{fig:epsart}The equivalence between the elements of
quantum repeating devices and quantum seals.}
\end{figure}

Consequently, there is a rigorous quantitative relationship between
the parameters $F$, $G$ and $\alpha $, $\beta $ describing the
quality of quantum repeating devices and quantum seals,
respectively. Since the input and output states $\left\vert \psi \right\rangle $ and $%
\left\vert \psi _{k}\right\rangle $ and the user's estimated signal
state $\left\vert \phi _{k}\right\rangle $\ of quantum repeating
devices are equivalent to the initial and final states of the system
$\Psi $ and Bob's decoded result $b_{i}$ of quantum seals,
respectively, by the definitions of the parameters $F$, $G$ and
$\alpha $, $\beta $, we yield
\begin{equation}
F=1-\beta   \label{e1}
\end{equation}%
and%
\begin{equation}
G=\alpha .  \label{e2}
\end{equation}

Nevertheless, we should note that the existing theories of quantum
repeating devices and quantum seals focus on different species of
information-disturbance tradeoff. On one hand, the existing theory
of quantum repeating devices generally studies merely the case where
the user wants to decode the \textit{quantum} aspect of the
information encoded in the quantum states. The case where the user
tries to decode only the \textit{classical} information encoded in
the quantum states was left out in literature. But as mentioned in
Sec. I, in many practical communication settings including QKD, an
essential problem is that an eavesdropper wants to know the
classical information only. That is, he needs not to distinguish the
quantum states exactly, as long as these states are corresponding to
the same classical information. Therefore the existing theory of
quantum repeating devices contributed less to the security analysis
of quantum
communication. On the other hand, the security of protocols sealing \textit{%
classical} information (either strings or a single bit) was well
studied in literature, while it still remains unclear how (in)secure
the protocols sealing \textit{quantum} information can be. For this
reason, our current finding on the equivalence between quantum
repeating devices and quantum seals is instructive. It indicates
that we can apply the existing theories interchangeably and find
interesting results for both fields. In the following, we will
present some examples.

\section{Fidelity bounds and optimality of quantum repeating devices for
classical information}

In this section, we will use the security bounds of quantum seals to
find interesting results on quantum repeating devices. Consider the
quantum repeating device decoding one single bit of classical mutual
information from quantum states. In this case, the bound (\ref{qs2})
applies. Using Eqs. (\ref{e1})
and (\ref{e2}), it can be rewritten as%
\begin{equation}
G-F\leq 1/8.  \label{qr5}
\end{equation}%
Note that in this case, $G$ is the estimation fidelity of the decoded \textit{%
classical} (instead of quantum) information. Therefore the bound (\ref{qr2})
obtained in the quantum case is not necessarily applied. Consequently, the
bounds (\ref{qr3},\ref{qr4}) no longer exist. Indeed, when $G-1/2>1/6$,
i.e., $G\geq 2/3$, we have%
\begin{equation}
(F-2/3)^{2}+4(G-1/2)^{2}\geq 4(G-1/2)^{2}>1/9.  \label{qr6}
\end{equation}%
This inequality holds for any dimension $d$ because Eq. (\ref{qs2})
and its derivative Eq. (\ref{qr5}) are valid regardless the
dimension of the Hilbert space of the input states. In the $d=2$
case, i.e., the input state is a qubit, Eq. (\ref{qr6}) clearly
shows that the quantum bound (\ref{qr4}) is surpassed for any
quantum repeating device which can decoded one classical bit of
mutual information from the input quantum states with the estimation
fidelity $G\geq 2/3$. Such a value of $G$ can indeed be reached in
real settings. For example, in the original BB84 QKD protocol
\cite{qi365}, the quantum states $\left\vert 0\right\rangle $ and
$\left\vert +\right\rangle $ both encode the classical bit $0$,
while the states $\left\vert 1\right\rangle $ and $\left\vert
-\right\rangle $ both encode $1$. Then a quantum repeating device
can be designed as follows. Measure the input states in the
Breidbart basis \cite{Breidbart}, i.e., $\{\cos (\pi /8)\left\vert
0\right\rangle +\sin (\pi /8)\left\vert 1\right\rangle ,\cos (5\pi
/8)\left\vert 0\right\rangle +\sin (5\pi /8)\left\vert
1\right\rangle
\}$, and output \textquotedblleft $0$\textquotedblright\ (\textquotedblleft $%
1$\textquotedblright ) if the measurement result is $\cos (\pi /8)\left\vert
0\right\rangle +\sin (\pi /8)\left\vert 1\right\rangle $\ ($\cos (5\pi
/8)\left\vert 0\right\rangle +\sin (5\pi /8)\left\vert 1\right\rangle $). In
this case, randomly distributed input bits can be decoded with the
estimation fidelity $G=\cos ^{2}(\pi /8)\simeq 0.8536>2/3$.

One of the significance of this result is that it indicates that the
security of quantum communication channel needs to be evaluated more
conservatively. This is because the transmission fidelity $F$, being the
measure on how well the input quantum state is preserved, is related
directly to the probability of detecting the eavesdropper who uses the
quantum repeating device to decode information on the state. Higher $F$
means less successful probability of the detection. Therefore Eq. (\ref{qr6}%
) surpassing Eq. (\ref{qr4}) means that $F$ can be higher for the
same $G$ if the eavesdropper decodes only the classical information
instead of trying to know the exact form of the input quantum state.
Such a case is exactly what the eavesdropper does in most quantum
communication we are interested today (e.g., QKD). Therefore Eq.
(\ref{qr5}) will be more appropriate than Eqs. (\ref{qr3},\ref{qr4})
for evaluating the security of such quantum communication channels.

Another question immediately followed is whether optimal quantum
repeating devices for decoding quantum information [i.e., whose $F$
and $G$ saturate the bounds (\ref{qr3},\ref{qr4}), for example, the
schemes proposed in Ref. \cite{qi586}] are still optimal for
decoding classical information. Before giving an answer, we must
notice what \textquotedblleft optimal\textquotedblright\ means in
the latter case. As shown above, when the purpose of the quantum
repeating device is to decode a classical bit
only, the quantum bound (\ref{qr2}) is gone. It was proven in Ref. \cite%
{impossibility} that perfect quantum seals can reach $\alpha =1$ and $%
\beta =0$ simultaneously. Therefore, using such quantum seals as the
schemes of quantum repeating devices can reach $F=G=1$. That is, if
by \textquotedblleft optimal\textquotedblright\ we want to favor all
users so that each of them can decode the bit with an estimation
fidelity as high as possible, while leaving the carrier as less
disturbed as possible, there exist perfect quantum repeating
devices. But in this case, the quantum communication channel is
trivial. This is because such perfect quantum repeating devices,
being a direct analog of perfect quantum seals, will then have to
encode different values of classical bits with orthogonal quantum
states \cite{impossibility}. Since no nonorthogonal states are
necessary, the channel can be completely classical. For example,
simply writing a bit on a piece of paper and passing it through all
the users can reach $F=G=1$.
In this sense, the quantum repeating devices saturating the bounds (\ref{qr3},%
\ref{qr4}) are surely not optimal for decoding classical information.

Here we consider another meaning of optimal. That is, our purpose is
changed into trying to saturate the bound (\ref{qr5}) to find a
balance between a high-estimation fidelity $G$ and a
low-transmission fidelity $F$. Since a lower $F$ means a higher
$\beta $, our purpose means to find the balance of detecting
eavesdroppers with a high probability, while still keeping the
encoded bit highly readable. [Note that we do not want the quantum
repeating devices that can saturate the bound (\ref{qs1}) because it
is indicates in Ref. \cite{He} that the input states must contain
zero amount of information of the encoded bits to saturate this
bound.] Optimal quantum repeating devices for decoding quantum
information are also not necessarily optimal for this purpose. This
is because, as shown above, the bound (\ref{qr5}) can violate the
bounds (\ref{qr3},\ref{qr4}) for certain values of $G$. On the other
hand, In Ref. \cite{He}, an optimal scheme of quantum seals that
saturates the bound (\ref{qs2}) was proposed. Namely, Alice should
seal the bit $b$ in the form
\begin{eqnarray}
\left\vert \phi _{b}\otimes \psi _{b}\right\rangle &=&\frac{\sqrt{3}}{2}%
\sum\limits_{i}c_{b,i}^{(b)}\left\vert \hat{f}_{i}^{(b)}\right\rangle
\left\vert \hat{e}_{i}^{(b)}\right\rangle  \nonumber \\
&&+\frac{1}{2}\sum\limits_{i}c_{b,i}^{(\bar{b})}\left\vert \hat{f}_{i}^{(%
\bar{b})}\right\rangle \left\vert \hat{e}_{i}^{(\bar{b})}\right\rangle .
\label{optimal}
\end{eqnarray}%
Here $\left\vert \hat{f}_{i}^{(b)}\right\rangle $'s ($\left\vert \hat{e}%
_{i}^{(b)}\right\rangle $'s) are the orthogonal states of Alice's system $%
\Phi $ (Bob's system $\Psi $) corresponding to the sealed bit $b$, with $%
c_{b,i}^{(b)}$'s being the superposition coefficients. This scheme has $%
\alpha =3/4$, $\beta =3/8$ thus reaches $\alpha +\beta =9/8$.
Therefore, optimal quantum repeating devices for decoding classical
information can be designed accordingly.

As a simplified example, suppose that in a quantum communication
channel, the classical bit $0$ is encoded as either
$(\sqrt{3}/2)\left\vert 0\right\rangle +(1/2)\left\vert
1\right\rangle $\ or $(\sqrt{3}/2)\left\vert 0\right\rangle
-(1/2)\left\vert 1\right\rangle $ and $1$ is encoded as either
$(1/2)\left\vert 0\right\rangle +(\sqrt{3}/2)\left\vert
1\right\rangle $\ or $(1/2)\left\vert 0\right\rangle
-(\sqrt{3}/2)\left\vert 1\right\rangle $. Then the optimal quantum
repeating device is to measure the input states in the basis
$\{\left\vert 0\right\rangle ,\left\vert 1\right\rangle \}$, and
output \textquotedblleft $0$\textquotedblright\
(\textquotedblleft $1$\textquotedblright ) if the measurement result is $%
\left\vert 0\right\rangle $\ ($\left\vert 1\right\rangle $). It can saturate
the bound (\ref{qr5}) for randomly distributed inputs.

Interestingly, the encoding method of the original BB84 QKD protocol
cannot allow quantum repeating devices saturating the bound
(\ref{qr5}). As calculated above, the best estimation fidelity is
$G=\cos ^{2}(\pi /8)$. According to Eq. (9) of Ref. \cite{He}, it
can be calculated that the transmission fidelity is
$F=1-2G(1-G)=3/4$, therefore we have $G-F\simeq 0.1036<1/8$\ in this
case. It will be interesting to study what advantages
can be brought when the optimal encoding method suggested by Eq. (\ref%
{optimal}) is adopted in QKD or other quantum cryptographic task.

\section{Security bounds for sealing quantum data}

Ever since the first proposal of the concept of quantum seals, it
became a major problem whether unconditionally secure quantum seals
exist. As reviewed in Sec. I, the security of the protocols sealing
classical data was already studied thoroughly \cite%
{impossibility,He,String,Security,qi505,qi516}. For the case where the
sealed information is quantum data, a protocol was proposed in Ref. \cite%
{Quantum}, but later found insecure by the author himself. However,
to date, there is still a lack of a general conclusion on the exact
security bound of the protocols sealing quantum data. Here, with the
equivalence between quantum seals and quantum repeating devices, we
can study the problem with the theory of the latter.

Before we proceed, it is important to note that it makes a major
difference whether the quantum data to be sealed is known to Alice
or not. If the quantum data are known to Alice, then the problem is
in fact equivalent to the sealing of classical information. Alice
can simply use a classical string to describe how to prepare a
quantum system whose state contains the quantum data to be sealed
and seal this classical string with QSS protocols such as the one
proposed in Ref. \cite{String}. With this method, even
sealing a single qubit can be secure. This is because for a qubit $%
\left\vert \psi \right\rangle =\cos \theta \left\vert 0\right\rangle +\sin
\theta \left\vert 1\right\rangle $, the value of $\theta $\ can have
infinite possibilities. It differs from the sealing of a single classical
bit, where the sealed bit $b$ can only have two possible values $0$ and $1$.
Therefore the insecurity proof of QBS \cite{He} does not apply to this case.
On the contrary, from the security proof of QSS \cite{String} it can be seen
that if Bob wants to decode $\theta $ with the reliability $\alpha
\rightarrow 1$, the probability for him to be detected will be $\beta
\rightarrow 1$ thus the protocol is secure.

Now let us focus on the case where the quantum data to be sealed are
unknown to Alice. To be rigorous, we further assume that Alice has
only one single copy of the quantum system containing these data, so
that her knowledge on the data is minimized. Let us apply the theory
of quantum repeating devices in this case. By combining Eqs.
(\ref{qr1}) and (\ref{e1}), we can see that any QS protocol using a
$d$-dimensional state to seal quantum data is bound
by%
\begin{equation}
\beta \leq 1-2/(d+1).  \label{qsqd}
\end{equation}%
It means that when the sealed data are a quantum state in a
high-dimensional Hilbert space ($d\rightarrow \infty $), the
existing theory provides little limitation on the detecting
probability $\beta $. Thus, a properly designed protocol may reach
$\beta \rightarrow 1$ when sealing a high-dimensional quantum state
and therefore can be regarded as secure. On the other hand, the
security level of QS for low-dimensional quantum states is bound
significantly by the dimensionality $d$. Especially, when $d=2$, we have%
\begin{equation}
\beta \leq 1/3.  \label{qsq}
\end{equation}%
Note that even when the sealed data are a qubit, the quantum state
used to seal it is not necessarily a two-dimensional state. If the
sealed qubit is mapped into a high-dimensional state and, most
important of all, if there is a method to force Bob to decode other
abundant information when he wants to decode the sealed qubit, then
QS protocols sealing a qubit can be made secure. Nevertheless, so
far we cannot prove the existence of this method. Therefore, before
such a method can be found in future researches, the security of QS
sealing a single qubit has to be bound by Eq. (\ref{qsq}). That is,
when Bob decoded the sealed qubit, Alice stands only $1/3$ chances
to detect it. Comparing to the security bound of QS sealing a single
classical bit \cite{He} (i.e., $\beta \leq 1/2$), sealing a qubit is
even less secure.

At last, we would like to note that Eqs. (\ref{qr3}) and (\ref{qr4})
cannot immediately give an analog of Eq. (\ref{qs2}) for QS
protocols sealing quantum data. This is because there is still a
subtle difference between QS and quantum repeating devices.
According to feature (2) of the above model of QS, Alice should
provide Bob with an operation $P$ for decoding. This operation can
generally enhance the reliability $\alpha $ of data decoded by Bob
(unless Alice wants to mislead Bob to the wrong outcome in the QS
protocol). On the other hand, in the quantum repeating device
problem, Eq. (\ref{qr2}) was obtained without assuming the existence
of
such a suggested operation. Therefore in QS protocols sealing quantum data, $%
G$ (i.e., $\alpha $) is not restricted by Eq. (\ref{qr2}) and
therefore the bounds in Eqs. (\ref{qr3}) and (\ref{qr4}) may be
surpassed. Exact bounds of $G$ will depend on the details of the
operation $P$ and vary for different protocols.

\section{Summary}

Thus it is shown that the transmission fidelity $F$\ and the
estimation fidelity $G$ describing the quality of quantum repeating
devices are related directly to the readability $\alpha $ and the
detecting probability $\beta $ of quantum seals. Therefore, the
theory in one field can be applied to the
other. With this method, we found that the existing fidelity bounds Eqs. (%
\ref{qr3},\ref{qr4}) of quantum repeating devices for decoding
quantum information can be surpassed when they are used for decoding
classical information from quantum states. Instead, the bound in the
latter case is Eq. (\ref{qr5}). Also, optimal quantum repeating
devices for quantum information are not necessarily optimal for
classical information. We also found that the security of protocols
sealing quantum data is bounded by Eqs. (\ref{qsqd},\ref{qsq}).

We thank Professor Matteo G. A. Paris for helpful discussions. The
work was supported in part by the NSF of China under Grant No.
10605041, the NSF of Guangdong province, and the Foundation of
Zhongshan University Advanced Research Center.

\end{document}